\documentclass{elsart}
\usepackage{graphics}
\usepackage{psfrag}
\usepackage{epsfig}
\usepackage{epsf}
\usepackage{float}
\usepackage{amssymb,stmaryrd,latexsym}
\usepackage{amscd}
\usepackage{mathrsfs}
\usepackage{color}
\newcommand{\be}{\begin{eqnarray}}
\newcommand{\ee}{\end{eqnarray}}
\newcommand{\beq}{\begin{equation}}
\newcommand{\eeq}{\end{equation}}

\definecolor{grey}{rgb}{0.3,0.3,0.3}

\usepackage{amsmath}
\usepackage{amssymb}
\usepackage{epsfig}

\newcommand{\spipi}{s_{\pi\pi}}

\begin{document}

\begin{frontmatter}
\title{Model-independent approach to $\boldmath{\boldsymbol\eta\to \pi^+\pi^-\gamma}$
and $\boldmath{\boldsymbol\eta^\prime\to \pi^+\pi^-\gamma}$}

\author{F.~Stollenwerk$^{1}$, C.~Hanhart$^{1}$, A.~Kupsc$^{2}$, U.-G.~Mei{\ss}ner$^{1,3}$ and A.~Wirzba$^{1}$}

{\small $^1$ Institut f\"{u}r Kernphysik (Theorie), Institute for Advanced
             Simulation and}\\
            {\small J\"ulich Center for Hadron Physics,}\\
            {\small Forschungszentrum J\"ulich,  D-52425 J\"{u}lich, Germany} \\
{\small $^2$ Department of Physics and Astronomy,}\\
           {\small Uppsala University,  Box 516, 75120 Uppsala, Sweden}\\
           {\small and High Energy Physics Department,}\\
          {\small The Andrzej Soltan Institute for Nuclear Studies,
          Hoza 69, 00--681 Warsaw, Poland}\\
{\small  $^3$ Helmholtz-Institut f\"ur Strahlen- und Kernphysik and} \\
         {\small   Bethe Center for Theoretical Physics,
%} \\ 
%{\small  
Universit\"at Bonn, D-53115
          Bonn, Germany} \\

\begin{abstract}
\noindent 
We present a new, model-independent method to analyze radiative decays of
mesons
to a vector, isovector pair of pions of invariant mass square below the first significant
$\pi\pi$ threshold in the vector channel. It is based on a combination of chiral perturbation
theory
and dispersion theory. This allows for a controlled inclusion of resonance
physics without the necessity to involve vector meson dominance explicitly.
 As an example, the method is applied to an analysis of the reactions $\eta\to
 \pi^+\pi^-\gamma$ and  $\eta^\prime\to
 \pi^+\pi^-\gamma$.
\end{abstract}

\end{frontmatter}

{\bf 1.} Radiative decays are known to be very sensitive tools to explore decay
mechanisms. Especially, when studied together with two hadrons as decay products, 
they enable us to adjust the invariant mass of the two-hadron system via 
a variation of the photon energy without interference of strong
three-body final state interactions. 

We here present a new, model-independent method to analyze the
mentioned radiative decays. The method is based on a combination of chiral perturbation
theory (ChPT) -- extended from $SU(3)$ to $U(3)$~\cite{HerreraSiklody:1996pm,Kaiser:2000gs} --  and a dispersive analysis. It is general and can be applied
to all decays of mesons with $\pi\pi\gamma$ final states, 
where soft bremsstrahlung does not occur and
where the pion pair is  of invariant mass square below the first {\em significant} $\pi\pi$ threshold, which is,  in
the  (iso)vector  case,  the $\omega\pi$ threshold.
In this work, however, we focus on the
decay of a pseudoscalar,  {\it i.e.} the $\eta$ and $\eta'$ meson, to a photon and a charged pion pair. In this
case the general selection rules enforce the pion pair to be in the isovector
channel. Especially for the decay of a relatively light meson, like the $\eta$, one
may expect that the ideal analysis tool is the effective field theory for 
the standard model at low energies, ChPT, and indeed the corresponding
calculation {has been} available to one-loop order for a long time~\cite{Bijnens:1989}. 
 But compared to modern data one observes a
significant deviation between the theory predictions
and data --- the source of which is mostly the non--perturbative $\pi\pi$
final state interaction. Several efforts have been made to include final state interactions by
unitarized extensions to the box-anomaly term -- the latter of course determines 
the $\eta\to\pi\pi\gamma$ decay in the
chiral limit.  The tree-level calculation can be enhanced by a
momentum dependent Vector Meson Dominance model \cite{Picciotto:1991} or more elaborate calculations
in the context of Hidden Local Symmetries \cite{Benayoun}. On top of the results at the
one-loop level an Omnes-function can be applied to describe the effects of $p$-wave pion scattering
\cite{Venugopal:1998fq,Holstein:2001} (see also \cite{Ko:1990xw}), or as done in the Chiral Unitary approach, a Bethe-Salpeter equation with
coupled channels can be used to generate resonances dynamically \cite{Borasoy:2004}.
We will later specify the details  which discriminate these mechanisms from our method.
The general problem with the vector meson dominance model is a priori that it
is unclear what relative strength {is to be} put between the tree level contribution
and the resonance contribution. This problem is resolved in the dispersion
theory approach as we will see below.

 For the transitions at hand it appears necessary to disentangle
perturbative and non--perturbative effects in a controlled way.
The method is
therefore split in two steps. In the {\em first step} the {spectral decay} data {are} fitted with
a function of the form (the details will be given below)
\begin{equation}
\label{eq:main}
\frac{d\Gamma}{ds_{\pi\pi}} = \left | A \,
P\left(s_{\pi\pi}\right)F_V\left(s_{\pi\pi}\right)\right|^2 \Gamma_0(s_{\pi\pi})\ ,
\end{equation}
where the normalization parameter $A$ has the dimension of ${\rm mass}^{-3}$ and where
\[
\Gamma_0(s_{\pi\pi}) = \frac{1}{3 \cdot 2^{11} \cdot \pi^3 m^3} \ 
\left(m^2-s_{\pi \pi} \right)^3 \ s_{\pi \pi} \ \sigma(s_{\pi\pi})^3
\]
collects phase-space terms and the kinematics of the absolute square of the
simplest gauge invariant matrix element (for point-particles).  The latter is
expressed through the $\pi\pi$--two-body phase-space
$\sigma(s_{\pi\pi})=\sqrt{1 - 4m_\pi^2/\spipi}$ in terms of the invariant mass
square $s_{\pi\pi}$ of the pion pair, while $m$ ($m_\pi$) denotes the mass of
the decaying particle (charged pion).  Since the initial state is a
pseudoscalar and the final state contains a photon, the partial wave of the
charged pion pair is expected to be dominated by $1^{--}$. If it can be
confirmed by experiment that the other partial waves can be neglected, then
the factorization given in Eq.\,(\ref{eq:main}) is exact and can be straight
forwardly derived using dispersion theory (see next section).  
If, however,  it eventually might turn out that higher partial waves would 
be needed to get a precise fit of  the angular data, then 
their contributions could still
be added in a perturbative way to our model-independent expression for the $p$-wave 
decay amplitude.

The pion vector
form factor $F_V(s_{\pi\pi})$ is known very well from both direct measurements
of $e^+e^-\to\pi^+\pi^-$~\cite{FF_exprefs,babarFF,KLOE,Na7FF} as well as
theoretical
studies~\cite{Gasser:1990bv,Bijnens:1998fm,FF_theoryrefs,Jose_new,yndurain,guerrero,ourff}.  It
collects all non-perturbative $\pi\pi$ interactions and is universal. On the
other hand, the function $P(s_{\pi\pi})$ as well as the normalization factor
$A$ are reaction specific and -- at least for the decay of light mesons or,
more accurately, for small values of $\spipi$ -- are expected to be
perturbative in the sense of ChPT.  In case of the $\eta$ and $\eta'$ decays
in the focus here, left-hand cut contributions are suppressed both
  kinematically, since the particle pairs in the $t$-channel are to be (at
  least) in a $p$--wave, and chirally, since the $p$--wave $\pi\eta$
  interaction starts at next--to--leading order only~\cite{BKM_91,bastian}. We may
therefore  expand $P(\spipi)$ in a Taylor series around $\spipi=0$ and define
\begin{equation}
P(s_{\pi\pi}) = 1 + \alpha s_{\pi\pi} + {\mathcal O}\left(s_{\pi\pi}^2\right)
\label{eq:poly}
\ .
\end{equation}
At higher order non-analytic terms, mainly from left-hand cuts, need to be considered. 
The parameters $\alpha$ and $A$ allow insights into the physics underlying the
decay process. Thus, in the {\em second step} of our method a proper matching scheme is
formulated, to relate the parameters $A$ and $\alpha$ to the parameters of the
underlying effective field theory. For the example at hand we will find that
this matching allows us, under certain assumptions, to impose a relation
between $\eta\to \pi^+\pi^-\gamma$ and $\eta^\prime\to \pi^+\pi^-\gamma$.

The paper is structured as follows: after a brief discussion of the pion
vector form factor, we apply the formalism to the mentioned
$\eta$ and $\eta'$ decays in Sec.\,3, extracting the phenomenological parameters
$\alpha^{(\prime)}$ and $A^{(\prime)}$. In Sec.\,4 those are then interpreted
via a matching to one--loop  $U(3)$ extended ChPT.  In Sec.\ 5 an interpretation of  essentially half of the empirical value of $\alpha^{(\prime)}$
is given, whereas Sec.\ 6 contains 
a comparison to earlier studies.
We close with a summary.

\begin{figure}[t]  % Figure 1
  \begin{center}
   \includegraphics[width=6.7cm]{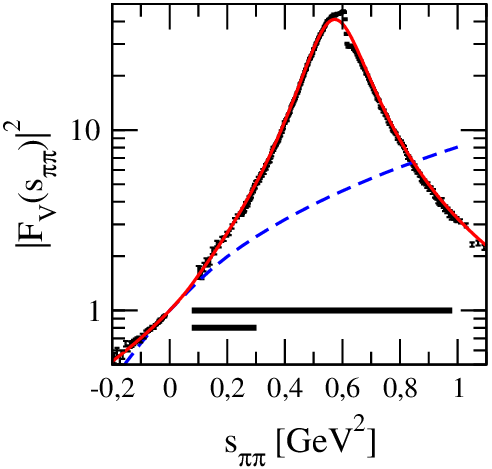}
   \includegraphics[width=6.7cm]{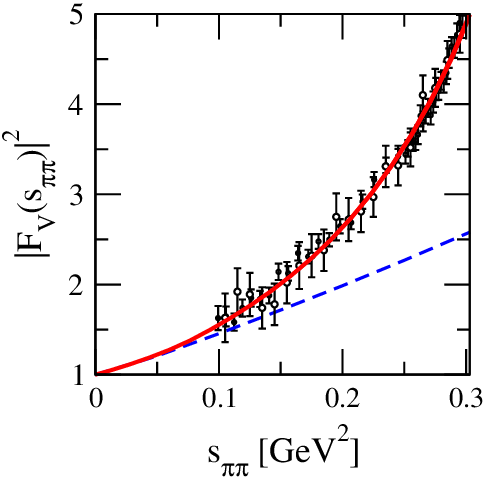}
   \caption{The (red) solid band shows the form factor derived from Eq.\,(\ref{eq:omnessub}), the
    (blue)  dashed line the result from one--loop ChPT (see Eq.\,(\ref{ffinchipt}))--- both with identical values for the
    pion radius. The time-like data from Refs.\,\cite{babarFF} and
    \,\cite{KLOE} are shown as solid and open circles, respectively. The
    space-like data
 are from Ref.\,\cite{Na7FF}.
    The short (long) thick, horizontal bar in the left panel denotes the kinematic range covered
    in the decay of the $\eta$ ($\eta'$). The right panel shows, as a linear
    plot,
a zoom into the $\spipi$ range relevant for the $\eta$ decay.
In this energy range the form factor can be approximated by the  polynomial
$|F_V(s_{\pi\pi})|\approx 1 + (2.12\pm 0.01) s_{\pi\pi}+ (2.13\pm 0.01) s_{\pi\pi}^2 + (13.80\pm 0.14) s_{\pi\pi}^3$ with $s_{\pi\pi}$ measured in units
of GeV$^2$. \label{fig:piFF}}
  \end{center}
\end{figure}

{\bf 2.} We start with a brief discussion of the pion vector form factor.
{In terms of the vector--isovector current $V_\mu^3$, i}t is defined via
\begin{equation}
\langle \pi^+(p')\pi^-(p)|V_\mu^3|0\rangle =  (p-p')_\mu F_V\left(s_{\pi\pi}\right) \ .
\end{equation}
In the elastic regime the form factor is, in symbolic notation, defined via
\begin{equation}
F_V = M_V + T_{\pi\pi} G_{\pi\pi} M_V \ ,
\label{eq:ffdef}
\end{equation}
with $M_V$, {$G_{\pi\pi}$} and {$T_{\pi\pi}$} for the production vertex, the two-pion propagator and
the $\pi\pi$ scattering amplitude, respectively. For simplicity, in this work
we
assume $M_V$ to be real which is exact for the transitions we will study
explicitly below.
Vector meson dominance models typically model the form factor, either, in its
simplest
variant
by a single term, $F_V(\spipi)=-m_\rho^2/(\spipi-m_\rho^2+im_\rho
\Gamma_\rho)$
{where $m_\rho$ and $\Gamma_\rho$ are the $\rho$ mass and width, respectively,} 
or by writing the second  term of Eq.\,(\ref{eq:ffdef})
as a (sum of) vector meson propagator(s) times $\spipi$, which leaves the relative strength
of the first and the second term as free parameter. We will here take a
different route which leaves no freedom of choice.
 From the  definition of Eq.\,(\ref{eq:ffdef}) it follows
 straight forwardly that
\begin{equation}
\label{eq:disc}
{\rm Im}(F_V(s_{\pi\pi})) = \sigma(s_{\pi\pi})\, T_{\pi\pi}(\spipi)^*\,F_V(\spipi) \ .
\end{equation}
 This relation holds for
the whole elastic regime which, in case of the pion vector form factor,
extends
to values of $\spipi$ well beyond 1\,GeV$^2$, although already at $\spipi = 16m_\pi^2$
formally the first inelasticity opens. Eq.\,(\ref{eq:disc}) is one way to
present the Watson theorem~\cite{book}: since Im$(F_V)$ is a real quantity, the phase of
the form factor has to agree {with} the phase of the
elastic scattering amplitude.
 In the elastic regime this equation
can be written as  ${\rm Im}(F)=\tan\delta_{11} {\rm Re}(F)$, with
$\delta_{11}$ for the elastic $\pi\pi$ phase shift in the vector channel. The
resulting
dispersion integral
 can
be solved analytically to give
\be
\label{eq:omnessub} F_V(s_{\pi\pi}) = \exp \left( {\textstyle \frac{1}{6} } s_{\pi\pi} \langle r^2
\rangle + \frac{s_{\pi\pi}^2}{\pi}
\int_{4m_\pi^2}^\infty ds
\frac{\delta_{11}(s)}{s^{2}(s-s_{\pi\pi}-i\epsilon)} \right) \ .
\ee
Contrary to   the standard procedure \cite{FF_theoryrefs}, 
we follow Refs.~\cite{guerrero,ourff} and use a twice subtracted dispersion
integral in order to guarantee that the integral over the phase  converges  in
the elastic regime.
The phase $\delta_{11}$ can be taken from
data or from theoretical analysis.
The phase used in the present analysis is taken from Ref.~\cite{Jose_new} (see
Eqs.\,(A7) and (A8) therein), valid up to $\sqrt{s_{\rm cut}}=1.4$ GeV --- we use a smooth
extrapolation of the phases beyond this energy following
Ref.~\cite{yndurain}. The phases agree with those of Ref.~\cite{royeq} up to
800\,MeV and with the available data.  Once the phase is fixed,
Eq.\,(\ref{eq:omnessub}) has only one free parameter --- the subtraction
constant $\langle r^2 \rangle$, the mean square charge radius of the pion.
Apart from the region around $\spipi = m_\omega^2$, where $\rho-\omega$ mixing
shows up very prominently, we find an excellent fit of the data for $\langle
r^2\rangle = 0.437(3)\,{\rm fm}^2$. The uncertainty includes the uncertainty
in the $\pi\pi$ phase shifts given in Ref.~\cite{Jose_new} as well as the weak
dependence on $s_{\rm cut}$. The form factor parametrization, including the
uncertainty, is shown as the red band in Fig.\,\ref{fig:piFF}.  The range for
the radius is consistent with $\langle r^2\rangle = 0.452(13)\,{\rm fm}^2$ of
Ref.\,\cite{BijnTala}  and $\langle r^2\rangle = 0.435(2)\,{\rm fm}^2$ of
Ref.~\cite{yndurain}.  The possible impact of $\rho-\omega$ mixing on the
$\eta'$ decay spectra is briefly discussed below.

As it provides the explicit solution for $F_V$ defined in
Eq.\,(\ref{eq:ffdef}), Eq.\,(\ref{eq:omnessub})  contain both the
Born term (pions going out without interaction) as well as the final state interaction.
In general, the mentioned dispersion integral fixes the form factor  up to
a multiplicative function that does not have right-hand cuts. Since the
right-hand discontinuity of the transition amplitude $\eta^{(\prime)}\to
 \pi^+\pi^-\gamma$ agrees with that of the pion vector form factor, 
the factorization employed in
Eq.\,(\ref{eq:main})  is  justified --- under the above-stated qualifications that the higher partial waves can be neglected and that  the left-hand cut contributions are suppressed.
 We chose the standard
normalization for the form factor, namely $F_V(0)$=1, which corresponds to
$M_V=1$ in Eq.\,(\ref{eq:ffdef}).

Remember that to one--loop order the expression for the form factor
reads~\cite{gasserleutwyler}
\be \label{ffinchipt}%
F_V(s_{\pi\pi}) = 1 + \frac{1}{6f_\pi^2}(s_{\pi\pi}-4m_\pi^2)\bar{J}(s_{\pi\pi})
+\frac{s_{\pi\pi}}{6} \left( \langle r^2 \rangle + \frac{1}{24 \pi^2 f_\pi^2} \right)\ ,
\ee
where the function $\bar {J}$ is defined in \cite{gasserleutwyler} and where $f_\pi = 92.2~{\rm MeV}$ denotes the pion decay constant. The small  kaon loop contribution is taken care of  by  the use of the empirical value of
the pion squared radius.

The full expression for the form factor,
Eq.\,(\ref{eq:omnessub}),
and its one--loop counterpart, Eq.\,(\ref{ffinchipt}),  which enters the 1-loop ChPT prediction for
 $\eta^{(\prime)} \to \pi^+\pi^-\gamma$ of  Ref.\,\cite{Bijnens:1989}\footnote{Actually, 
 Ref.\,\cite{Bijnens:1989} discusses the radiative two-pion decays of the octet and singlet states, $\eta_8$ and $\eta_1$.} 
are
compared 
with data in Fig.\,\ref{fig:piFF}. While by construction both curves correspond to the same
pion radius, the two descriptions start to
deviate already visibly at values of $\spipi$ as low as 0.09\,GeV$^2$.
The  kinematic range covered in the  radiative $\eta$ and $\eta'$ decays
spans from 0.077\,GeV$^2$ to 0.301\,GeV$^2$ and 0.918\,GeV$^2$, respectively.
The two ranges are indicated in the figure by the thick bars.
While one might hope to be able to describe the form factor in the
kinematic range for the $\eta$ within
ChPT at sufficiently high orders ({\it e.g.} below $0.25\,{\rm GeV}^2$ the 2-loop order  seems to be sufficient \cite{Gasser:1990bv,Bijnens:1998fm}), 
clearly a description for the $\eta'$
is impossible with any perturbative series.

As we will demonstrate below: once the non--perturbative part of the
transition amplitude in form of the pion vector form
factor is divided out, what remains can be described by a polynomial in
$\spipi$ of low order that can be analyzed within $U(3)$ extended ChPT, even in the case of the radiative
$\eta'$ decay.

\begin{figure}[t]   % Figure 2
	\centering
	\includegraphics[scale=0.58]{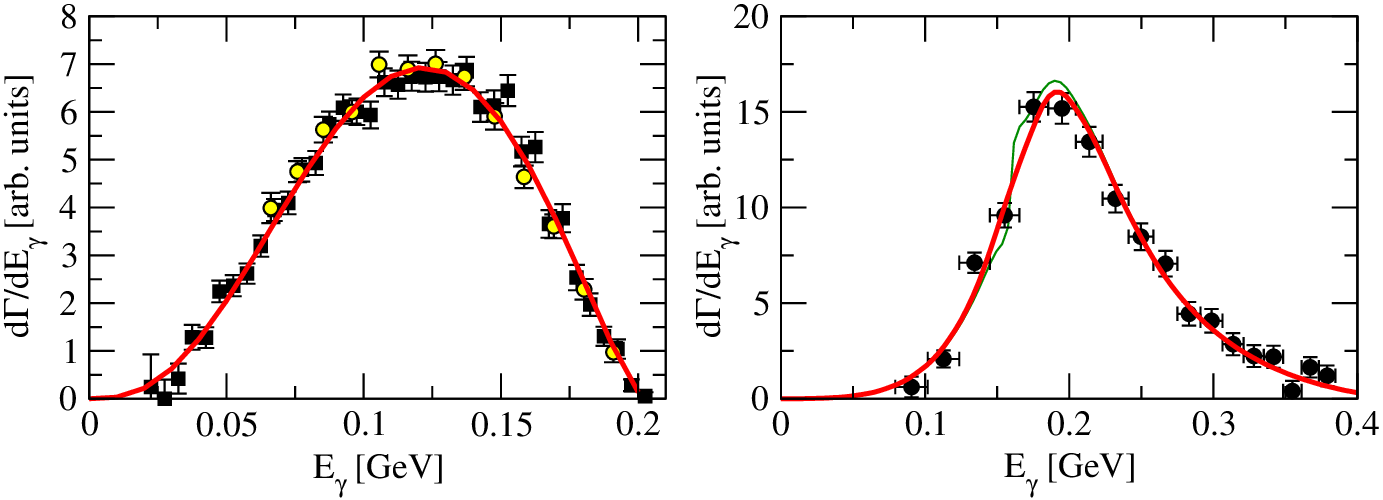}
	\caption{Experimental data and error weighted fits for $\eta$ (left, data are from Ref.\,\cite{wasa:2011}
	(filled squares) and Ref.\,\cite{Gormley:1970qz} (open circles)) and $\eta^\prime$ (right, 
	 data are from Ref.\,\cite{cry:1997}) to $\pi^+ \pi^- \gamma$
           according to Eqs.\,(\ref{eq:main}) and (\ref{eq:poly}) with 
            $s_{\pi\pi} = m_{\eta^{(\prime)}} ( m_{\eta^{(\prime)}}  -2
             E_\gamma)  $.
             The thin (green) line in the right panel denotes the possible impact of
             $\rho-\omega$ mixing under the assumption that it appears here with the same
           strength as in $F_V$.
	\label{fig:spectra}}
\end{figure}

{\bf 3.} We now turn to the evaluation of the full decay amplitudes.
As mentioned above the first step is to analyze the data for both the total decay rates as
well as the spectra with a fit of the parameters $A$ and $\alpha$ of Eq.\,(\ref{eq:main}).
Note that the experiments discussed below confirmed that the pion pair is in the
vector isovector channel.
The resulting fits to the spectra of the WASA-at-COSY collaboration \cite{wasa:2011} for the
$\eta$ case and of  the CRYSTAL BARREL collaboration \cite{cry:1997} for the $\eta'$ case  
are shown in Fig.\,\ref{fig:spectra}, left and right panel,
respectively. 
From the error weighted fits, we extract values of
\begin{equation}
 \alpha        = (1.96 \pm 0.27 \pm 0.02)~{\rm GeV}^{-2} \ ; \quad
 \alpha^\prime = (1.80 \pm 0.49 \pm 0.04)~{\rm GeV}^{-2} \label{eq:alphavalues}
\end{equation}
where the parameter extracted from the data on the $\eta'$ appears as primed.
The first and second uncertainty originate from the fit to the 
data on $\eta^{(')}\to \pi^+\pi^-\gamma$ and from that of the pion vector form
factor, respectively.
The $\alpha$ parameter was also determined directly by the WASA-at-COSY 
collaboration quoting in addition a systematic uncertainty of $0.59$ GeV$^{-2}$.
Since  the CRYSTAL BARREL data points include systematic
uncertainties, the uncertainty of the $\alpha'$ value should include
both statistical and systematic uncertainties.
We also studied other data sets for $\eta$ and $\eta^\prime$.
Concerning the former decay, Gormley {\it et al.}\,\cite{Gormley:1970qz} 
provides $\alpha = (1.8 \pm 0.4)~{\rm GeV}^{-2}$ while
Layter {\it et al.}\,\cite{Layter:1973ti} gives  $\alpha = (-0.9 \pm 0.1)~{\rm GeV}^{-2}$.
The acceptance correction of these old experiments was   derived  from the   
specified  
$d\Gamma/dE_\gamma$  distributions, respectively, under the assumption  that the pertinent matrix element is 
the  simplest gauge  invariant
one  (corresponding here  to   $P\left(s_{\pi\pi}\right)$  and
$F_V\left(s_{\pi\pi}\right)$ equal to one).
The Layter {\it et al.}
result seems to be
inconsistent both with WASA\,\cite{wasa:2011} and Gormley {\it et al.} \cite{Gormley:1970qz}.  However,  from the 
information provided
in  those old experimental papers it is
impossible to evaluate systematic uncertainties.
In case of the $\eta^\prime$, we obtain 
$\alpha^\prime = (2.7 \pm 1.0)~{\rm GeV}^{-2}$ from  the data of the GAMS-200 collaboration \cite{gams:1991}, which is larger,
but within error bars consistent with the value listed above. 
Hence, in the following, we use the values given in Eq. (\ref{eq:alphavalues}).

In the pion vector form factor $\rho-\omega$  mixing shows up as a quite
spectacular effect. In case of the
$\eta'$ decay, as a result of the $E_\gamma^3$ behavior of the rate at low
values of $E_\gamma$, the effect is a lot smaller: if we take the
mixing effect with the same strength as it appears in $F_V$ using the
prescription of Ref.~\cite{yndurain},  the impact of
the mixing on the $\eta'$ spectrum is very moderate --- see the wiggly (green) line
in the left panel of Fig.~\ref{fig:spectra}. A fit to the $\eta'$ spectrum
including the mixing as shown shifts the value of $\alpha'$ upwards by 0.3
GeV$^{-2}$, well within errors.

\begin{figure}  %Figure 3
	\centering
	\includegraphics[scale=0.58]{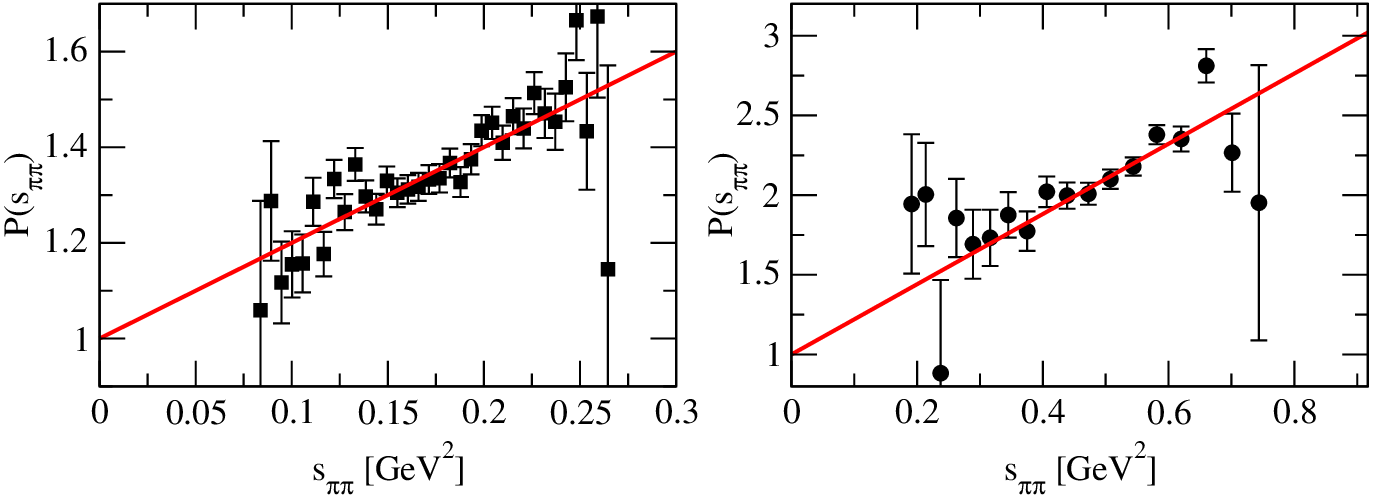}
	\caption{Plot of $P(s_{\pi\pi})$, defined in Eqs.\,(\ref{eq:main}) and (\ref{eq:poly}), as extracted from data
	for $\eta\to\pi^+\pi^-\gamma$ \cite{wasa:2011} (left panel) and $\eta'\to \pi^+\pi^-\gamma$ \cite{cry:1997}  
	(right panel).}
 	\label{fig:linplot}
\end{figure}

Instead of looking at the data themselves it is illustrative to extract from data directly the polynomial{s} $P(s_{\pi\pi})$.
These are shown for both radiative $\eta$ and $\eta'$ decays in the left and right panel of Fig.\,\ref{fig:linplot}, respectively.
Here one clearly sees that the residual $s_{\pi\pi}$ dependence for both transition amplitudes --- once the pion form factor and the phase space are divided out ---  has a linear behavior to a very good approximation. 
The statement is further corroborated by the fact that any {\em additional}
quadratic term to the linear polynomial with coefficients  as specified in Eq.\,(\ref{eq:alphavalues})
is compatible with zero:
$\beta= (0.21 \pm 0.67 \pm 0.06)~{\rm GeV}^{-4}$ and $\beta^\prime=(0.04 \pm 0.36 \pm 0.03)~{\rm GeV}^{-4}$.
This appears reassuring, although it came as a surprise that 
even for the $\eta'$ a first-order polynomial is sufficient.
The origin of this might be in the current quality of the data which is
best in the region of large values of $E_\gamma$ which corresponds to 
moderate values of $\spipi$ --- this is the region where the chiral expansion
is expected to converge (once resonance effects are taken out). This can also be
seen in Fig.\,\ref{fig:linplot}, right panel: clearly the fit is dominated by values of $s_{\pi\pi} \leq 0.6\,{\rm GeV}^2$ 
(this corresponds to pion relative momenta of at most $360\,{\rm MeV}$)  that are still reasonably smaller
than the typical hadronic scale of order of 1\,GeV, which sets the expected range of convergence for the
chiral expansion once the vector pion form factor is taken out.

The normalization of  the data used above cannot be employed to fix the
prefactor $A^{(\prime)}$,  for they are given in arbitrary units
only. However, once the slope parameters are fixed we can use the
experimentally measured partial widths \cite{pdg:2010}, $ \Gamma_{\eta
  \rightarrow \pi \pi \gamma}^{exp.}  = (59.8 \pm 3.8)\,{\rm eV}$ and $
\Gamma_{\eta^\prime \rightarrow \pi \pi \gamma}^{exp.} = (57.0 \pm 2.8)\,{\rm keV}$, to extract
\begin{equation}
 \delta        = -0.22 \pm 0.04 \ ; \quad
 \delta^\prime = -0.40 \pm 0.09 \label{eq:deltavalues} \ ,
\end{equation}
where we used the definition $A^{(\prime)}=A_0^{(\prime)}(1+\delta^{(\prime)})$,
with $A_0^{(\prime)}$ for the transition strength in the 
chiral limit\footnote{Here and in the following we will use
the term `chiral limit' in a somewhat loose sense, since we insert in $A_0$ the physical value of the
pion decay constant  $f_\pi$ and assume  that $f_8\neq f_\pi$ for the octet  pseudoscalar decay constant, etc.}, 
which is fixed for the transitions at hand by the chiral (box)
anomaly~\cite{Bijnens:1989,Holstein:2001}.

It should be stressed that the values extracted for $\alpha^{(')}$ are a
  lot more reliable at this stage than those extracted for $\delta^{(')}$,
  since the former are differential quantities while the latter are integrated
  ones, being quite sensitive to the shape of the spectrum also in the regime
  where we do not have high quality data yet. Thus, one might expect that,
  once these become available also for the $\eta'$ decay spectra at large
  values of $s_{\pi\pi}$, a linear polynomial is insufficient for
  $P(\spipi)$. This should not change $\alpha^{(')}$, however, it might
  significantly influence the integrated rate and therefore the values of
  $\delta^{(')}$.

{{\bf 4.}} In order to perform the matching {of the decay amplitude} to one--loop ChPT we now replace $F_V$ in
Eq.\,(\ref{eq:main}) by its one--loop expression, Eq.\,(\ref{ffinchipt}), and
expand the result to first order in $\spipi$. The resulting expression can 
then be equated with the corresponding one from one--loop ChPT, see appendix {\ref{app:expexp}}. This procedure
gives a relation between the phenomenological parameters $\alpha^{(\prime)}$
and
 $\delta^{(\prime)}$ and the $U(3)$ ChPT low-energy parameters 
 $a_1^{(8\pi)}$, $a_1^{(8K)}$, $a_1^{(1\pi)}$, $a_1^{(1K)}$ and $a_2$.
Especially one finds from the $\spipi$ independent terms for the $\eta$--decay
 \begin{flalign}
 \delta = \frac{1}{32 \pi^2 f^2_\pi} 
          & \biggl[ \frac{A_8}{A_0} \left (a_1^{(8\pi)} m_\pi^2 + a_1^{(8K)}  m_K^2\right)  
           +\frac{A_1}{A_0} \left(a_1^{(1\pi)}  m_\pi^2 + a_1^{(1K)}  m_K^2\right) 
    && \nonumber \\
   & \mbox{}- 4 m^2_{\pi} \log  { \frac{m^2_\pi}{\mu^2}}  
   + \frac{4 A_8 - 2A_1}{A_0} m_K^2
            \log  {\frac{m^2_K}{\mu^2}}  \biggr] &&
 \label{eq:delta_a1} 
  \end{flalign}
and from the $\spipi$ dependent terms for the $\eta$--decay
 \begin{flalign}
  \alpha +\frac16\langle r^2\rangle= \frac{1}{32 \pi^2 f^2_\pi}
    & \biggl[ a_2 - \frac 1 3 \log  {\frac{m^2_\pi}{\mu^2}}  -
    \frac{4A_8 + A_1}{6A_0} \log  {\frac{m^2_K}{\mu^2}}  && \nonumber \\ 
    & \hspace{4.cm}  \mbox{}- \frac{1}{9} \left( 8+  \frac{2A_8 - 7A_1}{2A_0} \right) \biggr] \  .&&
  \label{eq:alpha_a2}
  \end{flalign}
Here $m_K$ denotes the kaon mass. In line with previous investigations we
identify the scale $\mu$ with $m_\rho$. Furthermore we introduced the abbreviations
\[
A_8 = a \ \frac {e}{4 \sqrt{3} \pi^2 f^{3}_{\pi}} \frac {f_\pi}{f_8} \ ; \qquad
A_1 = b \ \frac {e\sqrt{2}}{4 \sqrt{3} \pi^2 f^{3}_{\pi}} \frac {f_\pi}{f_1} 
\] 
with $e$ the unit of electric charge, 
 $a=\cos (\theta)$ and $b=-\sin(\theta)$ in terms of  $\theta\simeq-20^\circ$
as the value  of the $\eta-\eta'$ mixing angle  and $f_8$ and $f_1$ the values of the octet
and singlet pseudoscalar decay constants -- all three values as specified as in
Refs.\,\cite{Venugopal:1998fq,Holstein:2001}, see also Ref.\,\cite{Bijnens:1989}. 
Since our aim  here is the comparison with the existing radiative decay results of the above mentioned references,  we still follow the old parameterization in terms of two (octet and singlet) decay constants and  {\em one} $\eta-\eta'$ mixing angle. This is in contrast to  more modern parameterizations
in terms of either two  octet-singlet decay constants and two mixing angles, which follow from the defining matrix elements of the octet
and singlet axial-vector current\,\cite{Leutwyler:1997yr}, or in terms of strange and non-strange decay constants  and  only one mixing angle, see {\it e.g.} Refs.\,\cite{Schechter,Feldmann:1998vh,Feldmann:1999uf}.

The transition strength in the chiral
limit is given by $A_0 = A_8\!+\!A_1$.  To arrive at the corresponding
expressions for the $\eta'$ decays, $\alpha$, $\delta$, $A_0$, $a$ and $b$
need to be replaced by their primed counter parts{, especially $a'=\sin (\theta)$ and
$b'=\cos(\theta)$}. In Eq.~(\ref{eq:alpha_a2}) we have only
kept the leading term of the kaon loop contributions, which turns out
to represent the complete expression to high accuracy.

In the case of the $\spipi$ independent terms,
we observe that  two parameters
from the phenomenological expression, $\delta$ and $\delta'$, have to be matched onto four parameters
from $U(3)$ extended ChPT, $a_1^{(8\pi)}$, $a_1^{(8K)}$, $a_1^{(1\pi)}$ and $a_1^{(1K)}$. Thus the matching does not provide any constraint. The situation is
different for the $\spipi$ dependent terms, as the two parameters $\alpha$
and $\alpha'$ have to be matched to the single parameter $a_2$. To justify
this
statement we needed to ignore $s_{\pi\pi}$ dependent counter terms subleading in $N_c$, the number of colors. Note,
however, that these neglected terms are used to cancel some of the additional divergences in the one-loop
approximation of the $\eta_1\to\pi^+\pi^-\gamma$ decay \cite{Bijnens:1989}.
Extracting $a_2$ from the $\eta$ and the $\eta'$ decays gives
\[
 a_2^\eta=9.70\pm 0.7 \ ; \qquad a_2^{\eta'} = 9.23 \pm 1.4 \ ,
\]
respectively. 
Thus we indeed find that the data sets for the $\eta$ and $\eta'$ 
decays are
consistent
with the assumption that the only  non--perturbative part of the amplitude
originates from the pion vector form factor.

{\bf 5.}  In the following, by invoking chiral Ward identity and large $N_c$ arguments,  we will give a physical interpretation to essentially half of the parameter $\alpha^{(')}$.
When the hidden gauge approach\,\cite{Bando:1984ej,Bando:1985rf} to
low-energy hadron physics was first applied to the anomalous sector, it was
already found out that the so-called {\em complete} VMD (for both the triangle
and the box anomaly sector) was unsustainable in this approach
\cite{Fujiwara:1984mp}, see also the reviews
\cite{Bando:1987br,Meissner:1988,Harada:2003jx}: while the non-anomalous and the
triangle-anomaly sectors were fully compatible with VMD, the description of
the box-anomaly-induced decays gave  satisfying predictions (see {\it
  e.g.}  the $\omega\to \pi\pi\pi$ decay)  only if the VMD triangle anomaly terms
were supported by a point-vertex (contact term) involving a photon and three pseudoscalars.
Cohen\,\cite{Cohen} was the first (see also \cite{Rudaz})
to point out that the chiral Ward identity implies that both the chiral triangle and the box anomaly
contribute to the $\gamma\pi\pi\pi$ processes and the $\eta^{(')}\to \pi\pi\gamma$ decays away from the chiral limit.
With the help of the low-energy theorem for the $\gamma\to \pi\pi\pi$ amplitude 
of Ref.\,\cite{Aviv:1971hq},  he argued that the total amplitude for these (in the chiral limit by the 
box-anomaly induced) processes
can be decomposed as 
 \begin{equation}
   {\cal A}^{\rm tot} ={\textstyle\frac{3}{2}} {\cal A}^{{\sf VVA}} - \half {\cal A}^{{\sf VAAA}}\ ,
  \label{eq:Cohen}
 \end{equation}
where the triangle or ${\sf VVA}$-type amplitude [${\sf V}$: vector, ${\sf
    A}$: axialvector],
under the assumption that   {\em one} of the vectors 
of the ${\sf VVA}$-type amplitude subsequently decays   into two pions, 
and the box or ${\sf VAAA}$-type amplitude 
contribute with a relative weight of ${\textstyle\frac{3}{2}}:- \half$. 
Still there is an ongoing debate whether these ${\sf VAAA}$-type
contact terms are empirically needed or not, see {\it e.g.} \cite{Truong,Benayoun}.

The vector pion form factor $F_V(s_{\pi\pi})$, which -- as mentioned before --  
contains both the Born terms and final state $\pi\pi$ interactions, 
can be represented by one of  the
${\sf V}$ legs in the triangle anomaly ${\sf VVA}$  vertex 
(the other ${\sf V}$ leg denotes the outgoing (isovector) photon, whereas the ${\sf A}$ leg stands for the decaying 
pseudoscalar $\eta$ under PCAC). 
In other words the ${\sf VVA}$ vertex
\begin{equation}
{\textstyle\frac{3}{2}} {\cal A}_{\eta\to\pi\pi\gamma}(0) F_V(s_{\pi\pi}) \tilde P(s_{\pi\pi}) 
\label{VAA}
\end{equation}
 (with 
$\tilde P(s_{\pi\pi})$ = $1 + \tilde \alpha s_{\pi\pi} + {\cal O}(s_{\pi\pi}^2)$  a polynomial, similar to $P(s_{\pi\pi})$ in
Eq.\,(\ref{eq:main}))
contains both {\em tree-leve}l contributions of the vector form factor, which are of leading order  in the $1/N_c$ expansion, and {\em genuine loop} 
contributions of $F_V$, which
are suppressed by at least one factor of  $1/N_c$ and are therefore subleading\footnote{Note that the prefactor 
${\cal A}_{\eta\to\pi\pi\gamma}(0)$ contains the coefficients $N_c/f_\pi^3 \sim {\cal O}(N_c^{-1/2})$  and therefore determines the leading overall $N_c$ 
scaling behavior of the
total  $\eta^{(')} \to \pi\pi \gamma$ decay amplitude. It  is of course the same for the ${\sf VVA}$ and the ${\sf VAAA}$ case.}.
 In the case of the  ${\sf VAAA}$ box anomaly, however,
 the two remaining ${\sf AA}$ legs stand for  two
 {\em separate} pions which either are already {\em the final pions} or which still {\em rescatter}. The first case,  which is the leading-$N_c$
contribution of the  ${\sf VAAA}$  process (times the ${\cal O}(N_c^{-1/2})$ scaling of the prefactor  ${\cal A}_{\eta\to\pi\pi\gamma}(0)$), would correspond  in the
 ${\sf VVA}$ scenario  just to the 
replacement  $F_V(s_{\pi\pi}) \to 1$, {\it i.e.} to the trivial term in
the Taylor-expansion of the vector form factor (\ref{eq:omnessub}),
while the  second (rescattering) case, since it necessarily involves an additional four-pion vertex, is  represented by the {\em subleading}  terms in the $1/N_c$ expansion
of   $F_V(s_{\pi\pi})$. 
The leading-order terms of the {\sf VVA}- and {\sf VAAA}-type 
amplitudes have to have the same $N_c$ scaling, 
since both are {\em tree-level} processes and since the initial state ($\eta^{(')}$) 
 and the final state ($\pi^+ \pi^- \gamma$) are the same, respectively.
 However, only the former  can run through a vector meson pole at tree-level.
The total ${\sf VAAA}$ result   is therefore
\begin{equation}
-{\textstyle\frac{1}{2}} {\cal A}_{\eta\to\pi\pi\gamma}(0) e^{- \frac{1}{6}\widetilde{\langle r^2\rangle}  s_{\pi\pi} } F_V(s_{\pi\pi}) 
 \tilde P(s_{\pi\pi}) \ .
 \label{VAAA}
\end{equation}
Here the coefficient $\widetilde{\langle r^2 \rangle}$ 
is  the leading $N_c^0$ contribution of the 
mean square charge radius of the pion, which to ${\cal O}(m_\pi^2)$ in ChPT is given by
$\langle r^2 \rangle= (\bar l_6 -1)/(16 \pi^2 f_\pi^2)$, 
see {\it e.g.} Ref.\,\cite{ourff}, such that
$\widetilde{\langle r^2 \rangle} = (\bar l_6 - \bar l_6^{\,N_c^0})/(16 \pi^2 f_\pi^2)$.
 Remember that $\bar l_6$ itself is of order $N_c$ and that $f_\pi$ scales as $\sqrt{N_c}$.
The exponential term $\exp(-\frac{1}{6} \widetilde{\langle r^2\rangle} s_{\pi\pi})$  in the ${\sf VAAA}$ vertex  therefore serves to remove
the leading {$N_c^0$} contributions in $F_V(s_{\pi\pi})$,  such that -- with exception of the trivial term -- only   the subleading ones are left over
(multiplying $-\frac{1}{2}  {\cal A}_{\eta\to\pi\pi\gamma}(0)$).
Summing the ${\sf VAA}$ and ${\sf VAAA}$ induced contributions 
to the $\eta^{(')}\to \pi\pi\gamma$ decay amplitude, {\it i.e.} (\ref{VAA}) and (\ref{VAAA}),  and Taylor-expanding the
exponential factor of (\ref{VAAA}) and $\tilde P(s_{\pi\pi})$, we therefore get the total amplitude
\begin{eqnarray}
{\cal A}_{\eta\to\pi\pi\gamma}(s_{\pi\pi}) =
 {\cal A}_{\eta\to\pi\pi\gamma}(0)
F_V(s_{\pi\pi}) \Bigl[ 1 + \underbrace{\left( {\textstyle\frac{1}{12}} \widetilde{\langle r^2\rangle} + \tilde \alpha\right)}_{\equiv \alpha} s_{\pi\pi} 
+{\cal O}(s_{\pi\pi}^2) \Bigr] \ .
\end{eqnarray} 
Thus, from this point of view, the coefficient $\alpha$ comprises of two terms:
 the leading $N_c^0$ term of the subtraction constant of the vector form factor,
$\half \widetilde {\langle r^2 \rangle}/6 \approx \bar l_6/(192 \pi^2 f_\pi^2) \approx 1.03\,{\rm GeV}^{-2}$ \cite{ourff}, 
 and the remainder $\tilde \alpha$. Comparing to the numbers of
 Eq.\,(\ref{eq:alphavalues}), we find in this case that about half of the value of
 $\alpha^{(\prime)}$ can be interpreted as coming from $\widetilde {\langle r^2 \rangle}$.

{\bf 6.} Finally, we will link our method to earlier studies and compare results with vector meson dominance considerations in general.
Inspired by an $N/D$ analysis of the process $\gamma\to \pi\pi\pi$ (see also
\cite{Ko:1990xw,Truong}), Holstein and Venugopal~\cite{Venugopal:1998fq,Holstein:2001}
constructed a precursor to the relation (\ref{eq:main}).  They started from an
ansatz containing a contact term as well as a rescattering term parameterizing
the $\pi\pi$ final state interactions, which reads in the notation used above
\begin{equation}
{\cal A}_{\eta\to\pi\pi\gamma}(s_{\pi\pi})= {\cal A}^{0}_{\eta\to\pi\pi\gamma} \left[ 1-c + c(1+a s_{\pi\pi})F_V(\spipi) \right]\ .
 \label{eq:Holstein}
\end{equation}
Here $c$ and $a$ are  free real-valued parameters 
and $ {\cal A}^{0}_{\eta\to\pi\pi\gamma}$ is the amplitude in the chiral limit.
By matching this ansatz  -- both at order  ${\cal O}(p^6)$ --  to one-loop chiral perturbation theory\,\cite{Bijnens:1989}  (more precisely,
 to  the coefficient in Ref.\,\cite{Bijnens:1989} of the standard one-loop function  $\bar J(s_{\pi\pi})$, \textit{cf.} Eq.\,(\ref{ffinchipt})) 
on the one hand and
to VMD (as in Ref.\,\cite{Picciotto:1991}) on the other hand, these free parameters were fixed  in Refs.\,\cite{Venugopal:1998fq,Holstein:2001} 
to $c=1$ via ChPT and to $a=1/(2 m_\rho^2)$ via VMD, respectively.
In light of the discussion presented above and as implicitly stated in Refs.\,\cite{Venugopal:1998fq,Holstein:2001}, 
$c=1$ ({\it i.e.} no contact term) appears as a necessity from
dispersion theory. Note that Ko and Truong\,\cite{Ko:1990xw} derived the same expression from unitarity and
the above-discussed Ward identity constraints.
If the parameter $a$ in Refs.\,\cite{Venugopal:1998fq,Holstein:2001} were not extracted from  VMD
or $\alpha'=0$ were not implicitly assumed as  in Ref.\,\cite{Ko:1990xw}
{\em and} the amplitude ${\cal A}_{\eta\to\pi\pi\gamma}(0)$ were not fixed to its chiral limit value $ {\cal A}^0_{\eta\to\pi\pi\gamma}$,
then the resulting relations 
of  Refs.\,\cite{Venugopal:1998fq,Holstein:2001} and of Ref.\,\cite{Ko:1990xw}
would have had the form of 
Eq.\,(\ref{eq:main}), for $P(s_{\pi\pi})$ expanded to first order.

Let us now compare  to vector meson dominance in general: in
its most simple form one would just get $\alpha=0$, at variance with data, see (\ref{eq:alphavalues}).
If, following Ref.~\cite{Cohen},
 the chiral
Ward identities are implemented, then the simplest scenario would  correspond to  $\tilde \alpha=0$  instead of  $\alpha=0$,
a result at the edge of being consistent with current data.
This is 
what was used in Refs.~\cite{Venugopal:1998fq,Holstein:2001,Ko:1990xw}. 
Finally note that in all   
evaluations that  modify  the $\eta\to\pi^+\pi^-\gamma$ decay amplitude  of the chiral limit 
solely by $s_{\pi\pi}$ dependent terms, {as is the case} in vector meson dominance models, 
it is impossible to simultaneously predict the experimentally
measured shape of the distribution and the 
empirical 
branching ratio.

{\bf 7.} To summarize,   the $\eta^{(')}\to\pi^+\pi^-\gamma$ decay amplitude, here assumed to be $p$-wave dominated,
factorizes into a universal
part and a reaction specific part after the trivial (point-particle) kinematics is 
removed. If higher partial waves cannot be neglected, their contribution should be added  suitably 
to the above-mentioned $p$-wave amplitude.
The universal part, which applies to all $p$-wave dominated
radiative decays with an isovector $\pi^+\pi^-$ pion pair in the final state,
is given by the well established vector pion form factor $F_V$.  The reaction
specific part can be parametrized as an expansion in the invariant mass
$s_{\pi\pi}$ of the pion pair: $P\left(s_{\pi\pi}\right)$ times the
normalization factor $A$ --- as long as left-hand cut contributions are suppressed, as is the case for the reactions under consideration.
The expansion can  then be treated perturbatively, since unitarity and
analyticity of the final-state interaction dictate that the pion form factor
already takes care of  the non-perturbative aspects of the $\pi\pi$
unitary cut. Moreover, the perturbative expansion allows a systematic
comparison with ChPT predictions.  Especially, the case with $\eta'$ decays
shows a way to extend methodically perturbative calculations also to the
resonance region.
In this sense the presented approach is indeed {\em model-independent}. 

For the description of the present world data it is
sufficient to consider only the linear term (the $\alpha^{(')}$ parameter) 
in the
expansion of$P\left(s_{\pi\pi}\right) = 1 + \alpha^{(')} s_{\pi\pi}+ {\cal O}(s_{\pi\pi}^2)$ --- both for the $\eta$ and $\eta'$
decays.  
The extracted values of this parameter for the existing
experiments  are $\alpha~=(1.96 \pm 0.27_{\rm stat} \pm 1.00_{\rm syst})~{\rm GeV}^{-2}$ and $\alpha^\prime~= (1.80 \pm 0.49_{\rm tot})~{\rm GeV}^{-2}$.
The data  of   Gormley  {\it et al.} \cite{Gormley:1970qz} and 
the new data of the  WASA-at-COSY collaboration \cite{wasa:2011} are
consistent (both support an $\alpha$ parameter of the order  $2\,{\rm GeV}^{-2}$),  
whereas  the data of  Layter  {\it et al.} \cite{Layter:1973ti}  indicate  an  $\alpha$  value  below
zero. However,  the old  experiments (Gormley and  Layter) do  not provide
estimates of the systematic uncertainties.

We propose to use the introduced $\alpha$ and $\alpha'$ coefficients or, more generally, the polynomial $P(s_{\pi\pi})$ to
parametrize and compare -- including the pertinent statistical and
systematical uncertainties -- the spectra of all past and future radiative
decay experiments  with only one (iso)vector p-wave $\pi^+\pi^-$ pair in the final state.  
New   experimental  data   on   the  $\eta\to\pi^+\pi^-\gamma$   decay
distributions  from KLOE  are  at  the final  stages  of the  analysis
\cite{Ambrosino:2011gn}  and will  be released  soon.   Meanwhile also
WASA-at-COSY,  CLAS  and  BES-III  have  collected  new  large  data  sets  of
respectively $\eta$  and $\eta'$  decays that will  enable us to  check if
further terms in the expansion of $P(s_{\pi\pi})$ are necessary.

\noindent 
{\bf Acknowledgements}

We are grateful to  Bastian Kubis for useful discussions and would like to thank
Christoph  F. Redmer and the WASA-at-COSY collaboration and
Camilla Di Donato and the KLOE collaboration for  providing information about
the experimental  data.  This work  was in part  supported by
the Helmholtz Association through funds provided to the Virtual Institute
``Spin and strong QCD'' (VH-VI-231), by the DFG (TR 16, ``Subnuclear Structure of Matter''),
by the EU HadronPhysics2 project ``Study of strongly interacting matter'' and by
the European Commission under  the 7th Framework Programme through the
``Research  Infrastructures''  action  of  the  ``Capacities''  Programme:
FP7-INFRASTRUCTURES-2008-1, Grant Agreement No.\,227431.

\appendix

\section[Appendix]{Explicit expressions from one--loop  ${\boldmath\boldsymbol{U(3)}}$ extended ChPT 
 \label{app:expexp}}

The ChPT amplitude employed in the matching {(\textit{cf.} Eqs. (\ref{eq:delta_a1}, \ref{eq:alpha_a2}))} has been obtained as follows. 
The divergences occurring in the regularized one--loop expression \cite{Bijnens:1989} are absorbed by anomalous $\mathcal O(p^6)$ counter terms --
a complete list of which is presented in \cite{Bijnens:2001bb,Ebertshauser:2001nj} in the $SU(3)$ case. We 
follow Ref.\,\cite{Borasoy:2004}  where also some $U(3)$ extensions 
of these terms can be found.   
However, we neglect the pure singlet terms containing  the coefficients $\bar W_{13}$ and $\bar W_{14}$ of Ref. \cite{Borasoy:2004}, which posses an $s_{\pi\pi}$-dependence, but have one more trace than their $SU(3)$ counter parts, such that they are only subleading in the $1/N_c$ expansion. Furthermore, we
use the Gell-Mann--Okubo formulae 
to dispose the $\eta$ \cite{GMOeta} and $\eta^\prime$ \cite{GMOetaprime} masses,  which is correct to ${\cal O}(p^6)$.  In
summary, we get the ChPT amplitude
\begin{equation}
 {{\cal A}}^{\rm ChPT} = \big[ {A_8} \cdot C_{\eta_8} + {A_1} \cdot C_{\eta_1} \big]
                  \epsilon_{\mu \nu \alpha \beta} \ (\epsilon_{\gamma}^\star)^\mu
                       (p_\gamma)^\nu (p_+)^\alpha (p_-)^\beta  
\label{eq:Achpt_r}
\end{equation}
with 
\begin{eqnarray}
 C_{\eta_8} &=& 1 + C_{\eta_8}^{\rm loops} + \frac{1}{32 \pi^2 f^2_\pi} \big[ a_1^{(8\pi)}  m^2_\pi + a_1^{(8K)}  m^2_K + a_2  s_{\pi \pi} \big] \ ,                  \label{eq:Achpt_r_eta8} \\                          
 C_{\eta_1} &=& 1 + C_{\eta_1}^{\rm loops} + \frac{1}{32 \pi^2 f^2_\pi} \big[ a_1^{(1\pi)}  m^2_\pi + a_1^{(1K)}  m^2_K + a_2  s_{\pi \pi} \big] \ .   \label{eq:Achpt_r_eta1} 
\end{eqnarray}
Here the coefficients $C_{\eta_i}^{\rm loops}$ comprehend the finite loop contributions (see \cite{Bijnens:1989}), while 
the remaining constants comprise the  low-energy constants (LECs) of Ref.\,\cite{Borasoy:2004}  in the following way:
\begin{eqnarray}
 a_1^{(8\pi)} &=& \kappa \cdot \left[ -4 \bar w^{(0)}_7 - 8 \bar w^{(0)}_8 - {\textstyle\frac{7}{3}} \bar w^{(0)}_{11} - 2 \bar w^{(0)}_{12} \right] \ , \\
 a_1^{(8K)}   &=& \kappa \cdot \left[ 8 \bar w^{(0)}_8 + {\textstyle\frac{4}{3}} \bar w^{(0)}_{11} \right] \ , \\
 a_1^{(1\pi)} &=& \kappa \cdot \left[ -4 \bar w^{(0)}_7 + 2 \bar w^{(0)}_8 +12 \bar w^{(0)}_9 + 3 \bar w^{(0)}_{10} - 5 \bar w^{(0)}_{11} - 2 \bar w^{(0)}_{12} \right] \ , \\
 a_1^{(1K)}   &=& \kappa \cdot \left[ 4 \bar w^{(0)}_8 + 6 \bar w^{(0)}_{10} + 4 \bar w^{(0)}_{11} \right] \ , \\
 a_2        &=& \kappa \cdot \left[ 2 \bar w^{(0)}_{11} + \bar w^{(0)}_{12} \right]   
\end{eqnarray}
with $\kappa = 2^{11} \pi^4 f_\pi^2$. 
Hence the momentum dependent terms of $\eta_8$ and $\eta_1$ involve only \emph{one} linear combination of 
the LECs (\textit{cf.} $a_2$), whereas the mass terms are governed by \emph{four} linear combinations.

Note that  the structure of $C_{\eta_8}$ and $C_{\eta_1}$ displayed in Eqs.\,(\ref{eq:Achpt_r_eta8}) and (\ref{eq:Achpt_r_eta1}) 
can also be derived from \cite{Bijnens:2001bb,Ebertshauser:2001nj}, if the Gell-Mann--Okubo formulae and the
pertinent $U(3)$ extension are applied as above.

\end{document}